# A theoretical study on the performances of thermoelectric heat engine and refrigerator with two-dimensional electron reservoirs

Xiaoguang Luo,[1,*] Jizhou He,[2] Kailin Long,[1] Jun Wang,[1] Nian Liu,[3] and Teng Qiu[1,*]

[1]*Department of Physics, Southeast University, Nanjing 211189, China*

[2]*Department of Physics, Nanchang University, Nanchang 330031, China*

[3]*Department of Physical and Electronics, Anhui Science and Technology University, Bengbu 233100, China*

## Abstract

Theoretical thermoelectric nanophysics models of low-dimensional electronic heat engine and refrigerator devices, comprising two-dimensional hot and cold reservoirs and an interconnecting filtered electron transport mechanism have been established. The models were used to numerically simulate and evaluate the thermoelectric performance and energy conversion efficiencies of these low-dimensional devices, based on three different types of electron transport momentum-dependent filters, referred to herein as: $k_x$, $k_y$ and $k_r$ filters. Assuming the Fermi-Dirac distribution of electrons, expressions for key thermoelectric performance parameters were derived for the resonant transport processes, in which the transmission of electrons has been approximated as a Lorentzian resonance function. Optimizations were carried out and the corresponding optimized design parameters have been determined, including but not limited to the universal theoretical upper bound of the efficiency at maximum power for heat engines, and the maximum coefficient of performance for refrigerators. From the results, it was determined that $k_r$ filter delivers the best thermoelectric performance, followed by the $k_x$ filter, and then the $k_y$ filter. For refrigerators with any one of three filters, an optimum range for the full width at half maximum of the transport resonance was found to be $< 2k_B T$.

---

[*] Email address: 276718626@qq.com (X. Luo); tqiu@seu.edu.cn (T. Qiu).

I. **Introduction**

In recent years, the study of thermoelectric materials has attracted significant interest, due to their potential benefits in enhancing the efficiency of thermal to electric conversion processes by reducing the levels of wasted heat energy and hence the demands on electricity and natural energy resource utilization [1]. Through thermoelectric Seebeck and Paltier effects, forward and backward energy conversion processes that employ solid state materials are now becoming plausible without the use of moving parts, enabling greener, cleaner, more efficient thermoelectric conversion technologies. To evaluate the effectiveness of thermoelectric materials, the dimensionless thermoelectric figure of merit, $ZT = S^2 \sigma T / (k_e + k_l)$, is usually adopted, where $S$ is the Seebeck coefficient (also known as thermal power), $\sigma$ is the electric conductivity, $T$ is the average temperature of the material (with hot/cold junctions, $T_{H/C}$), and $k_{e/l}$ is thermal conductivity due to the electron/lattice. The figure of merit is closely related to the conversion efficiency, which can be improved by making $ZT$ as large as possible. Generally, bulk materials suffer from a very low figure of merit. Values of $ZT$ for bulk materials around unity (corresponding to the conversion efficiency of <10%) cannot meet the requirements of commercial applications [2]. How to enhance $ZT$ has therefore become an important research topic. With developments in nanoscaled device technology, it is now evident that the use of smaller-sized thermoelectric materials may offer solutions to the present bottleneck for achieving higher values of $ZT$. Examples of such materials include: $Bi_2Te_3$/$Sb_2Te_3$ superlattice nanomaterial [3], silicon nanowires [4], PbSeTe-based quantum dots [5], and other complex nanoscaled compounds [6]. Very few methods for achieving values of $ZT > 3.0$ currently exist, although somewhat larger values have been reported in the literature for certain materials. From the given expression for the figure of merit, it is immediately apparent that increased values of $ZT$ can be achieved by: (1) increasing the Seebeck coefficient; (2) increasing electric conductivity; and/or (3) reducing thermal conductivity. Some years ago, Hick et al. [7] proposed that the reduced dimensionality of superlattices could be used to enhance the electronic density of states and to obtain increased efficiency. Indeed, the heat loss due to phonons in low-dimensional nanomaterials can be effectively reduced by employing materials with a scale-size approaching the phonon mean-free path. Other research suggests that low-dimensional nanostructures with delta-shaped transport distributions may represent excellent thermoelectric materials [8], especially in semiconductor applications [9]. Theoretically, the thermoelectric process using such materials would appear to be reversible, offering a corresponding figure of merit $ZT \to \infty$, where the potential for high energy conversion efficiencies close to the maximum theoretical Carnot limits of $\eta_C = 1 - \tau$ for the efficiency of power generation, and $\varepsilon_C = \tau/(1-\tau)$ for coefficient of performance (COP) of cooling, where $\tau = T_C / T_H$ [10].

It should be noted that larger efficiency does not always mean better for energy conversion. Taking the heat engine as an example, the power output usually decreases with the increasing efficiency, and may even tend to zero when the efficiency achieves the Carnot limit for the engine, when strong coupling exists between energy and matter flow. Hence, optimization is required to achieve meaningful heat engine designs, and a commonly preferred design criteria is one that delivers best efficiency at maximum power (EMP) $\eta_{\max P}$. Historically, the most famous EMP limit is the Curzon-Ahlhorn (CA) efficiency: $\eta_{CA} = 1 - \sqrt{\tau}$, which is obtained in an endoreversible Carnot heat engine by using finite-time thermodynamic theory [11]. Actually, CA efficiency is just the upper bound of EMP for low symmetric dissipation Carnot engine [12], a linear irreversible thermodynamic system operating under conditions of strong coupling between the energy and matter flow. And it can be surpassed in several different kinds of nonlinear irreversible systems, see the excellent review by Tu [13] and the references therein. Taking the different statistics of the working substance as instances, both Maxwell-Boltzmann [14] and Fermi-Dirac [15] statistics have been shown produce a larger EMP than $\eta_{CA}$ in strong coupling engines, and the maximum EMP for Bose-Einstein substance is nearly equal to $\eta_{CA}$. The region defined by: $\eta_C / 2 < \eta_{\max P} < \eta_C / (2 - \eta_C)$ has been confirmed to bound these EMP limits mentioned above, although it is also obtained from low-dissipation Carnot engines [16,17]. So it is plausible that the maximum EMP of the thermoelectric heat engine should also be bound by this region.

Most current theoretical models of thermoelectric devices consist of low-dimensional hot and cold reservoirs with low-dimensional electronic conductors. For example, the zero-dimensional (0D) quantum dot between two one-dimensional (1D) reservoirs (i.e., the 1D-0D-1D system) can effectively enhance the thermoelectric performance through resonant tunneling or energy level hopping [18-21], due to the delta-shaped transport distribution of the electronic conductor.

Actually, enhanced thermoelectric performance can be optimized through the transmission function [22], by using conductors that are designed as momentum-dependent (or energy-dependent) filters. Low-dimensional conductors between two three-dimensional (3D) reservoirs can also result in high conversion efficiencies, as reported in [23,24]. In this previous research, electrons in the free direction were modeled with Maxwell-Boltzmann statistics for which the average energy component of electrons in one free-direction is simply $k_\mathrm{B}T/2$, where $k_\mathrm{B}$ is the Boltzmann constant. However, it is considered that those treatments are too simplistic; the energy component is still very dependent on the Fermi-Dirac distribution of electrons, and the expression should be renormalized to describe the electron transmission in low-dimensional conducting systems. To assist in developing a more generalized analytical model for nanoscaled thermoelectric devices, the development of detailed theoretical models for two-dimensional (2D) hot and cold reservoirs, with three kinds of interconnecting electron transport material designed to function as momentum-dependent filters (electron conductors), has been adopted as the focus for our current research. These three filters include: (i) a $k_x$ filter to constrain electrons in the direction of transport, (ii) a $k_y$ filter to constrain electrons in the direction perpendicular to transport, and (iii) a $k_r$ filter to constrain electrons in all directions. Many 2D nanomaterials such as graphene, MoS$_2$, topological insulators have become popular in recent material science and physics research. The thermoelectric properties of these 2D materials are attractive in terms of potential new technology applications [25-27], and the further study we have embarked on is therefore considered to be worthwhile.

In the research presented herein, we have adopted classical 2D electron systems as the hot and cold reservoirs (e.g., such those confined in the well of a GaAs/AlGaAs interface) and Lorentzian resonant transport as the filtering mechanism for the nanoscaled thermoelectric devices in our study. This model was chosen because it facilitates, with a reasonable level of complexity, the theoretical analysis and expressions for the thermoelectric parameters and performance of the nanoscaled heat engines and refrigerators. Moreover, using numerical computation, detailed optimizations can be carried out in relation to each device's power output, cooling rate, energy conversion efficiency, and figure of merit. To the best of our knowledge, this is the first time that a theoretical model has been developed, which derives the maximum $\tau$-dependent EMP of a thermoelectric heat engine with 2D reservoirs. In the remaining sections of this article: Sec. II describes the model and analyzes the transport behavior of 2D electrons; Sec. III derives expressions for several thermoelectric performance parameters for different momentum filters; Sec. IV discusses these important performance parameters and their detailed optimization in detail and Sec. V presents the main conclusions of our research and summarizes key results.

**II. Thermoelectric devices consist of 2D reservoirs and momentum filters**

Two 2D electronic reservoirs (one hot, one cold) connected with an electronic conductor form the model of our nanoscaled thermoelectric device. The cold one possess a higher electrochemical potential compared to its hot counterpart, so that the electrons can be exchanged due to the temperature and electrochemical potential gradients, where $T_H > T_C$ and $\mu_C > \mu_H$. Heat flux coupled with the exchanged electrons can then be transferred or converted between the reservoirs by means of the interconnecting medium. It is noted that electrochemical potentials are related to the voltage applied to the reservoirs. However, it can still be assumed that the electron energy $E$ is measured relative to the same voltage-independent zero (e.g., the bottom of the conduction band for a semiconductor) as the electrochemical potentials, when $\mu_C - \mu_H \ll \mu_{C/H}$. The electron-exchange behavior can also be explained by that the electrons are often transmitted from the occupied states to the empty states. The difference between Fermi-Dirac distribution functions of the hot and cold reservoirs ( $f_{H/C} = 1/\{1+\exp[(E-\mu_{H/C})/k_B T_{H/C}]\}$ ) can be observed in the inset of Fig. 1a, where $f_C = f_H$ when the electron energy $E = E_0 = (\mu_C T_H - \mu_H T_C)/(T_H - T_C)$, and $f_{C/H} > f_{H/C}$ when $E$ is lesser or greater than $E_0$. Additionally, when $E < E_0$, electron flow can occur from cold to hot reservoir, while for $E > E_0$, the electron flow will be in the reverse direction. Combined with the change of heat $E - \mu_{H/C}$ associated with the electron transmission, the energy spectrum of the electrons in the thermoelectric device can be segmented into three regions [10,24], comprising: cooling region when $\mu_C < E < E_0$, power generating region when $E > E_0$, and thermal heating region when $0 < E < \mu_C$ where the cold reservoir is simply heated by thermal exchange. The energy distribution across these three regions is what largely determines whether the thermoelectric device works as a refrigerator or a heat engine.

Electrons in these reservoirs can also be regarded as free electrons at effective mass assumption. Ignoring other edge effects, two cases have been considered in the studies presented below, including frameworks of parallel interfaces of reservoirs with an infinite longitudinal dimension, and one reservoir encircling the other concentrically. Due to the quantum confinement or barrier blockage as illustrated in Fig. 1a-1c, at least three kinds of filtering mechanism are applicable to these thermoelectric devices, referred to herein as: $k_x$, $k_y$ and $k_r$ filters (where $k_r^2 = k_x^2 + k_y^2$), due to their different directionality characteristics. Since we are dealing with 2D reservoirs, $k_{x/y}$ filtered and $k_r$ filtered cases can be regarded as 2D-1D-2D and 2D-0D-2D systems, respectively. All three kinds of filters can be realized in the parallel case (refer to Figs. 1a-1c), while in the encircled case, only the $k_r$ filter appears plausible (Fig. 1d).

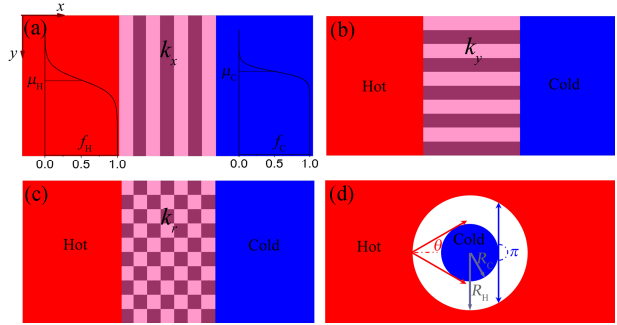

**Fig. 1.** (Color online) Schematics of three electron transport momentum-dependent filters. (a) $k_x$ filter, e.g., superlattices alternated parallel to the interface of the 2D reservoirs; (b) $k_y$ filter, e.g., superlattices alternated perpendicular to the interface of the 2D reservoirs; (c) $k_r$ filters, e.g., superlattices alternated in two directions or quantum dots directly connected to the 2D reservoirs. (d) Schematic of the encircled case with cold inner part (radius, $R_C$) and hot outer part (inner radius, $R_H$), where $\theta$ is the effective half-scattering angle of electrons in the hot reservoir, and $\pi$ is the effective scattering angle of electrons in the cold reservoir. The insets in (a) are the corresponding Fermi-Dirac distributions of the hot and cold reservoirs, where $\mu_H < \mu_C$.

In our nanoscaled thermoelectric devices, not all electrons in the reservoirs are available to participate in exchange processes. Many factors such as the Fermi-Dirac distribution, the transmission behavior in the electronic conductor, and even the geometry of the reservoir interfaces [28], may impact on the electron-exchange process dramatically. The electron transmission in two-reservoir system should be dependent on both of reservoirs and conductor, and the number of electrons flowing out of a 2D reservoir per unit time can be expressed as:

$$\dot{N}_{H/C} = \frac{2}{h^2} \iint L_{H/C} v_x f_{H/C} t \, dp_x dp_y \tag{1}$$

where the factor of "2" accounts for the degeneracy of electron spin, $h$ is the Planck constant, $L_{H/C}$ is the edge length of the hot/cold reservoir interface, $v_x = \hbar k_x / m^*$ is the group velocity of electrons in x-direction where $\hbar = h/2\pi$, and $m^*$ is the effective mass of electrons in the reservoir. Under the parabolic approximation, momentum $p$ and energy $E$ are: $p_{x/y} = \hbar k_{x/y}$, $E = (\hbar k)^2 / 2m^*$. The integral conditions and the transmission probability $t$ are very dependent on the properties of the reservoir interfaces and the electronic conductor, and the latter can filter electrons selectively based on their momentum vectors. Moreover, $\dot{N}_{H/C}$ is not real in practice, but it can induce the measurable net electronic flux, represented by: $\dot{N} = \dot{N}_H - \dot{N}_C$.

## III. Heat flux formulas for different electron transport filters

For the parallel case with $L_H = L_C$, all the thermoelectric parameters can be obtained per unit length per unit time with no need to consider the entire edge length of the 2D reservoirs. Coupled with the net electronic flux (i.e., $\dot{N} = \dot{N}_H - \dot{N}_C$), an amount of heat $E - \mu_{H/C}$ will be released or absorbed by a reservoir when an electron leaves or arrives via the electron-exchange medium. The current ($I$) and heat flux out of the hot/cold reservoir ($\dot{Q}_{H/C}$), per unit length per unit time, can be expressed by the following equations:

$$I = \frac{2e}{h^2} \int_{-\infty}^{\infty} \int_0^{\infty} v_x (f_H - f_C) t(E_i) \, dp_x dp_y \tag{2}$$

$$\dot{Q}_{H/C} = \frac{2}{h^2} \int_{-\infty}^{\infty} \int_0^{\infty} (E - \mu_{H/C}) v_x (f_{H/C} - f_{C/H}) t(E_i) \, dp_x dp_y \tag{3}$$

where $e$ indicates the charge of an electron. These equations are valid for all of $k_x$, $k_y$ and $k_r$ filter parallel cases. Furthermore, when the resonant transport has one resonant peak of the first band, the transmission probability can be approximated well by a Lorentzian function [29], as:

$$t(E_i) = \frac{(\Gamma_i / 2)^2}{(E_i - E_i^p)^2 + (\Gamma_i / 2)^2} \tag{4}$$

where $i = x, y, r$ for the different filters, $E_r = E$, and $\Gamma_i$ and $E_i^p$ are the equivalent full width at half maximum (FWHM) and peak energy level at the center of resonance, respectively, both of which can be adjusted by the electrical gating applied on the electronic conductor.

When the thermoelectric device functions as a heat engine, an equivalent load path should form, with a current flowing against a finite potential of $\Delta V$, where $e\Delta V = \mu_C - \mu_H$. The power output and the efficiency may then be written as: $P \equiv I\Delta V = \dot{Q}_H + \dot{Q}_C$ and $\eta = P/\dot{Q}_H$, respectively. Similarly, once the heat flux flows out of the cold reservoir, the device can function as a refrigerator, for which the cooling rate and COP are $R = \dot{Q}_C$ and $\varepsilon = R/(-P)$, respectively. Additionally, the total entropy production rate of these thermoelectric systems is given by:

$$S = -\frac{\dot{Q}_H}{T_H} - \frac{\dot{Q}_C}{T_C}, \qquad (5)$$

which is valid for all $k_x$, $k_y$ and $k_r$ filters. From Eqs. (3)-(5), it can be demonstrated that: $S \geq 0$.

**A. Modeling the $k_r$ filter mechanism**

For this case, the $k_r$ filter mechanism confines the transmitted electrons in all the directions, as illustrated in Fig. 1c, where the working electrons are filtered according to their total momentum $\hbar k_r$. $k_r$ filter is equivalent to a total energy filter, since $E = (\hbar k_r)^2 / 2m^*$. The encircled case identified in Fig. 1d can be readily studied for the $k_r$ filter. For this case, the length of the reservoir interface is $L_{H/C} = 2\pi R_{H/C}$. If the inner reservoir is cold, as shown in Fig. 1d, all electrons from the cold reservoir can arrive at the hot reservoir, however, reciprocity does not apply to any reverse flow. For hot to cold reservoir electron flow, only the electrons located in the scattering angle $2\theta$ are effective in enabling the thermoelectric effect, where $\sin\theta = R_C / R_H$. In this instance, Eq. (1) can be rewritten in the following forms for the two flow directions:

$$\dot{N}_H^r = \frac{2L_H}{m^*h^2}\int_0^\infty \int_{-\theta}^{\theta} p_r^2 \cos\theta' f_H d\theta' t dp_r = \frac{4\sqrt{2m^*}L_C}{h^2}\int_0^\infty \sqrt{E} f_H t(E) dE \qquad (6a)$$

$$\dot{N}_C^r = \frac{2L_C}{m^*h^2}\int_0^\infty \int_{-\pi/2}^{\pi/2} p_r^2 \cos\theta' f_C d\theta' t dp_r = \frac{4\sqrt{2m^*}L_C}{h^2}\int_0^\infty \sqrt{E} f_C t(E) dE \qquad (6b)$$

Eq. (6a) implies a most interesting result that the electronic flux out of the hot outer reservoir is very dependent on the interface length of the inner cold reservoir $L_C$, and not $L_H$. Therefore, Eqs. (2) and (3) are still valid for the encircled case with a $k_r$ filter, per unit length of the inner reservoir interface. The heat flux flowing out of the reservoirs with a $k_r$ filter can then be written

as:

$$\dot{Q}_{H/C}^{r} = \int_{0}^{\infty} \psi_{H/C}^{r}(E) t(E) dE \quad (7a)$$

$$\psi_{H/C}^{r}(E) = \frac{4\sqrt{2m^{*}}}{h^{2}} (E - \mu_{H/C}) \sqrt{E} (f_{H/C} - f_{C/H}) \quad (7b)$$

**B. Modelling the $k_x$ filter and $k_y$ filter mechanisms**

In the cases applicable to $k_x$ and $k_y$ filters, the structures (e.g., superlattices) modeled herein are configured to filter electrons based on the momentum (or energy) component in only one direction, for which two orthogonal cases ($x$ and $y$) are considered. Transmitted electrons in $k_x$ and $k_y$ filter mechanisms, such as shown in Fig. 1a and 1b, are confined to the $x$-direction and $y$-direction, respectively. In the other direction for each filtering case, the electrons are considered to be free, and this results in an unavoidable thermal conduction that reduces the efficiency of energy conversion. Therefore, the corresponding 2D Schrödinger's equations for electron transport for each filter case can be separated into transverse and longitudinal parts. For convenience, O'Dwyer et al. [24] assumed that electrons in the "free" direction follow Maxwell-Boltzmann statistics, for which the average energy component for each degree of freedom can be easily calculated as: $k_B T / 2$. However, these assumptions are somewhat over-simplified, and our calculations start from the fundamental thermodynamic equations, following the Fermi-Dirac statistics of electrons.

For the parallel case with a $k_x$ filter, the electronic flux out of the reservoir can be calculated by Eq. (3), and rewritten in the form of Eq. (7) when the energy $E$ and function $\psi_{H/C}^{r}(E)$ are replaced by the $x$-component energy ($E_x$) and $\psi_{H/C}^{x}(E_x)$, respectively, yielding the result given below:

$$\psi_{H/C}^{x}(E_x) = \frac{\sqrt{2\pi m^{*}}}{h^{2}} \left\{ \begin{array}{l} 2(E_x - \mu_{H/C}) \left\{ \sqrt{k_B T_{C/H}} \text{Li}_{1/2} \left[ -e^{-(E_x - \mu_{C/H})/k_B T_{C/H}} \right] - \sqrt{k_B T_{H/C}} \text{Li}_{1/2} \left[ -e^{-(E_x - \mu_{H/C})/k_B T_{H/C}} \right] \right\} \\ + (k_B T_{C/H})^{3/2} \text{Li}_{3/2} \left[ -e^{-(E_x - \mu_{C/H})/k_B T_{C/H}} \right] - (k_B T_{H/C})^{3/2} \text{Li}_{3/2} \left[ -e^{-(E_x - \mu_{H/C})/k_B T_{H/C}} \right] \end{array} \right\},$$

(8)

where $\text{Li}_n(x) = \sum_{i=1}^{\infty} \frac{x^i}{i^n}$ is the polylogarithm function (also known as Jonquière's function).

Similarly, the $k_y$ filter can be realized in the orthogonal parallel case, as shown in Fig. 1b. The electrons in the electronic conductor are then confined in the $y$-direction and half-free in $x$-direction (e.g., $k_x > 0$). Therefore, from Eq. (3), the expressions of heat flux flowing out of the reservoirs also have the form of Eq. (7) when the energy $E$ and function $\psi_{H/C}^r(E)$ are replaced by energy component $E_y$ and $\psi_{H/C}^y(E_y)$, respectively, yielding the result given below:

$$\psi_{H/C}^y(E_y) = \frac{2\sqrt{2m^*/E_y}}{h^2}\left\{\begin{array}{l}(E_y - \mu_{H/C})\left\{k_B T_{H/C}\log\left[1+e^{-(E_y-\mu_{H/C})/k_B T_{H/C}}\right] - k_B T_{C/H}\log\left[1+e^{-(E_y-\mu_{C/H})/k_B T_{C/H}}\right]\right\}\\ -(k_B T_{H/C})^2 \text{Li}_2\left[-e^{-(E_y-\mu_{H/C})/k_B T_{H/C}}\right] + (k_B T_{C/H})^2 \text{Li}_2\left[-e^{-(E_y-\mu_{C/H})/k_B T_{C/H}}\right]\end{array}\right\}.$$

(9)

**IV. Performances and optimization analysis of thermoelectric heat engines and refrigerators with different filters**

**A. $\Gamma_i \to \infty$ (no filtering case)**

When $\Gamma_i \to \infty$, the thermoelectric devices perform as if there were no filtering, i.e., $t(E_i) = 1$ for all transmitted electrons. Actually, all three filters produce equivalent "no filtering" results in this instance. Even so, thermoelectric devices under these conditions can still work as heat engines or refrigerators, at least in theory. This can be demonstrated by computing the relative power output and cooling rate for such a device from Eq (3). The following equations apply:

$$P^* = P / \frac{\sqrt{2\pi m^*}}{h^2}(k_B T_H)^{5/2} = 2(x_H - \tau x_C)\left[\text{Li}_{3/2}(-e^{x_H}) - \tau^{3/2}\text{Li}_{3/2}(-e^{x_C})\right] \quad (10a)$$

$$R^* = \dot{Q}_C / \frac{\sqrt{2\pi m^*}}{h^2}(k_B T_H)^{5/2} = 2\tau^{5/2}x_C\text{Li}_{3/2}(-e^{x_C}) - 3\tau^{5/2}\text{Li}_{5/2}(-e^{x_C}) - 2\tau x_C\text{Li}_{3/2}(-e^{x_H}) + 3\text{Li}_{5/2}(-e^{x_H})$$

(10b)

where the dimensionless scaled electrochemical potential $x_{H/C} = \mu_{H/C}/k_B T_{H/C}$, and $\tau = T_C/T_H$ is as defined earlier. Thus, at a given hot reservoir temperature of $T_H$, the working regions for this thermoelectric device can be determined by $P^* > 0$ and $R^* > 0$. From numerical calculation for the no-filtering case (refer to Fig. 2), the working region as a heat engine extends initially and then shrinks gradually for increased values of $\tau$. When $\tau \to 1$, both boundaries of the working region tend to the line, $x_H = x_C$, and the region vanishes. However, the thermoelectric device works as a refrigerator only when $\tau > 0.691226$. The cooling region with respect to $x_{H/C}$ is shaped like a partially twisted leaf that appears to stretch when $\tau$ is increased. As $\tau \to 1$, the maximum $x_H$ of the cooling region tends to infinity, and the boundary close to the heat engine working region approaches the line, $x_H = x_C$, from above.

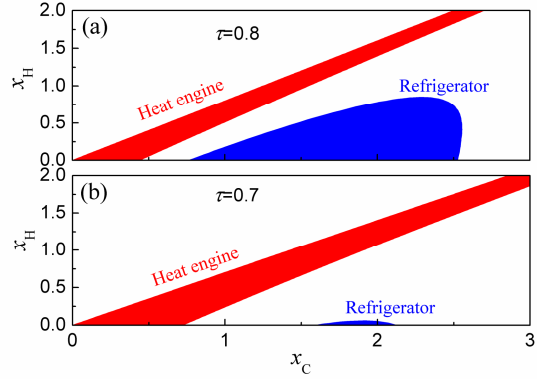

**Fig. 2.** (Color online) The working region of a heat engine (red) and a refrigerator (blue) at $\tau = 0.7$ and $\tau = 0.8$ when $\Gamma_i \to \infty$; the remaining (uncolored) thermal regions indicate where the cold reservoir is being heated simply by the heat flux from the hot reservoir.

These working conditions occur only under ideal hypothetical situations, and cannot be achieved optimally in practice. As already mentioned, the energy spectrum of the electrons in the thermoelectric device can be divided into three regions by the energy positions of $\mu_C$ and $E_0$. The transmitted electrons with energy $\mu_C < E < E_0$ and $E > E_0$ result in cooling and power generating effects, respectively. Therefore, filtering the electrons in the exchange process within these regions will be more practical and available.

**B. $\Gamma_i \ll k_B T$**

Nanostructures such as quantum dots and superlattices can be used to implement and fine-tune the various energy and momentum-dependent filtering mechanisms, with different operating parameters. The presence of transmission spectra with a single resonance peak provides for a simple and effective filter when the subbands are far away from the electrochemical potentials, around which by several $k_B T$ the electron-exchange occurs. As the FWHM of the resonance decreases to a very small value, fewer electrons are transmitted by the electronic conductor due to the Coulomb blockage. This behavior can therefore provide an effective means of improving the efficiency of nanoscaled thermoelectric devices [10,15]. For the extreme filtering cases considered here, where $\Gamma_i \ll k_B T$, the heat fluxes out of the reservoirs (refer to Eq. (3)) and the total entropy production rate of the system (refer to Eq. (5)) can be approximated, respectively, as:

$$\dot{Q}_{H/C}^i = \frac{\pi \Gamma_i}{2} \psi_{H/C}^i \left( E_i^p \right) \quad (11)$$

and

$$S^i = -\frac{\pi \Gamma_i}{2} \left[ \frac{\psi_H^i \left( E_i^p \right)}{T_H} + \frac{\psi_C^i \left( E_i^p \right)}{T_C} \right] \quad (12)$$

where $i = x, y, r$. The heat fluxes are very weak because of the small FWHM, while the efficiencies of the thermoelectric devices in this case are adequate. Similar to 1D-0D-1D and 3D-0D-3D systems, for the $k_r$ filter case, the efficiency for the heat engine mode and the COP for the refrigerator mode are expected as $\eta^r = \frac{\mu_C - \mu_H}{E_r^p - \mu_H}$ and $\varepsilon^r = \frac{E_r^p - \mu_C}{\mu_C - \mu_H}$, respectively. They are the maximum values, for variable FWHM, at specified levels of electrochemical potential [10,24]. Since $E_r^p$ can be adjusted by means of the electronic conductor, both the efficiency and COP tend towards the Carnot values when $E_r^p \to E_0$, as shown in Fig. 3, and are not dependent on the reservoir temperatures directly. Meanwhile, the entropy production rate vanishes and the system becomes reversible. The same situation does not happen with $k_x$ and $k_y$ filters. Owing to the degree of freedom in these cases, the entropy always increases until electron-exchange between the reservoirs cuts off at high energy. Regarding efficiency or COP values plotted in Fig. 3a for the stated parameter values, higher maximum values are generally achieved for $k_x$ filter compared to its $k_y$ counterpart. Compared to the results for 3D reservoirs based on Maxwell-Boltzmann analysis [24], different statistics of electrons and different dimensionalities of reservoirs in $k_r$ filter case cannot change the maximum energy conversion efficiencies (efficiency and COP). However, having analyzed this issue in some depth, we believe that the statistics and dimensionality assumptions do influence the maximum energy conversion efficiencies for both $k_x$ and $k_y$ filters.

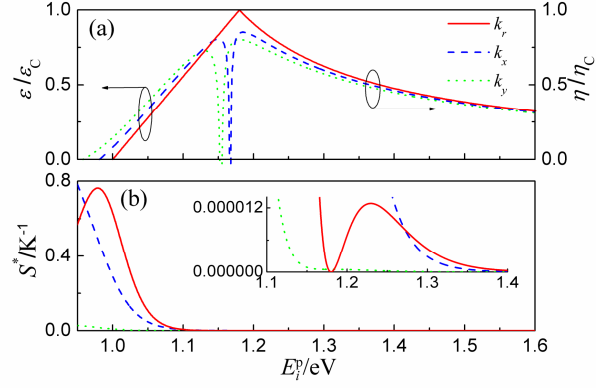

**Fig. 3.** (Color online) Performance properties of thermoelectric devices when $\Gamma_i \ll k_B T$. (a) Relative COPs (left axis) and relative efficiencies (right axis) of thermoelectric devices working as refrigerators and heat engines respectively. (b) The relative entropy production rate of the system $S^* = S/\Gamma$ (SI), (the inset provides partial enlarged detail). The parameters used are: $\mu_C = 1\text{eV}$, $\mu_H = 0.98\text{eV}$, $T_C = 270\text{K}$, $T_H = 300\text{K}$ and $m^* = 0.067m$ for GaAs, where $m$ is effective mass of electrons in vacuum,

**Table 1.** Different thermoelectric contributions of the electrons in different energy regions, parameters are the same as those shown in Fig. 3.

| Filters | Refrigerator (eV) | Heat engine (eV) |
|---|---|---|
| $k_r$ | $1 < E < 1.18$ | $E > 1.18$ |
| $k_x$ | $0.98113 < E_x < 1.16605$ | $E_x > 1.16774$ |
| $k_y$ | $0.957653 < E_y < 1.15207$ | $E_y > 1.15548$ |

Table 1 illustrates some interesting results based on the data of Fig. 3a. For $k_r$ filter, the thermoelectric device works as a refrigerator when $\mu_C < E < E_0$ and as a heat engine when $E > E_0$. When $E < \mu_C$, only thermal heating exchange occurs from the hot to cold reservoir. Similar situations occur with $k_x$ and $k_y$ filters. However, an additional very small thermal heating region exists between working regions of the refrigerator and heat engine at: $1.16605\text{eV} < E_x < 1.16774\text{eV}$ for $k_x$ filters, and $1.15207\text{eV} < E_y < 1.15548\text{eV}$ for $k_y$ filters. This is quite different compared to the $k_r$ filter case where no such gap appears. The abnormal gap is believed to be attributable to the degree of freedom in the "free" direction for both the $k_x$ and $k_y$ filters. During the electron-exchange process for these filters, the heat brought out of the cold reservoir by the net electronic flux cannot offset the intrinsic heat leakage (due to the temperature gradient) flowing from the hot to cold reservoir. Additionally, it seems that $k_y$ filter leaks more heat.

To reveal further insights into the physics of these devices, comparisons between the maximum efficiencies are discussed bellow. From Eqs. (8), (9) and (11), the efficiency and COP of $k_x$ and $k_y$ filters are quite dependent on the temperature, which implies the temperature-dependence of the figure of merit, $ZT$. The reason for this may be due to the unavoidable thermal conductivity. Taking the heat engine as an example, an upper bound for the efficiency can be obtained when $\kappa = (\mu_C - \mu_H)/k_B(T_H - T_C) \gg 1$. In this instance, the maximum efficiency is obtained when $E_i^p = E_0$ [24], which may be the reason why maximum efficiencies appear around $E_0$ in Fig. 3a. The factor $\kappa$ introduced here could be used, potentially, to measure the contributions of electrochemical potentials and temperatures for $k_x$ and $k_y$ filters. When $\kappa \gg 1$, temperatures would have a relatively small impact, resulting in relatively small unavoidable thermal conductivity. The analytical expressions for maximum efficiencies for $k_x$ and $k_y$ filtered heat engines can then be given by

$$\eta^x = \eta_C \left[ 1 + \eta_C \left( k_B T_C + k_B T_H + k_B \sqrt{T_C T_H} \right) \Lambda(\kappa) / 2eV_0 \right]^{-1} \qquad (13a)$$

$$\eta^y = \eta_C \left[ 1 - \eta_C \left( k_B T_C + k_B T_H \right) \Omega(\kappa) / eV_0 \right]^{-1} \qquad (13b)$$

where $\Lambda(\kappa) = \text{Li}_{3/2}(-e^{-\kappa})/\text{Li}_{1/2}(-e^{-\kappa})$ and $\Omega(\kappa) = \text{Li}_2(-e^{-\kappa})/\log(1+e^{-\kappa})$ tend to +1 and -1, respectively, when $\kappa \gg 1$.

From these equations, it can be shown that the upper bounds of maximum efficiency $\eta^x > \eta^y$ for any given preset bias of $V_0$, since $T_C + T_H > \sqrt{T_C T_H}$, and both bounds are smaller than the Carnot value. Therefore, compared to $k_y$ filter, $k_x$ filter case would appear to be more suitable for heat engine applications. The method used here cannot be adopted to deal with the refrigerators, because $\dot{Q}_C^{x/y} < 0$ when $E_{x/y}^p \to E_0$. However, detailed simulations (not shown here) have confirmed that the maximum COP for $k_x$ filtered refrigerator is larger than that of their $k_y$ counterparts. So, when the $k_r$ filters are difficult to be realized, $k_x$ filters may provide an alternative way to obtain higher efficiency.

In addition to the maximum efficiency, the power output is also a very important parameter for heat engines. However, maximum power is quite often achieved under conditions that are somewhat different to those required for maximum efficiency, and a similar phenomenon happens in the case of the refrigerator. In order to determine optimized working conditions, a target function of $\chi = \zeta \dot{Q}_{out}$ is usually used. Here $\zeta$ denotes efficiency or COP and $\dot{Q}_{out}$ denotes the heat flux flowing out of the target reservoir, e.g., $\chi = \eta \dot{Q}_H$ for a heat engine and $\chi = \varepsilon \dot{Q}_C$ for a refrigerator [30]. By maximizing this target function, the $\tau$-dependent upper bounds can be found, such as the CA efficiency $\eta_{CA} = 1 - \sqrt{\tau}$, and CA COP $\varepsilon_{CA} = 1/\left(1 - \sqrt{1-\tau}\right) - 1$, for a low symmetric dissipation Carnot engine and refrigerator, respectively [12]. Many related studies have used these approaches [31-34].

For a heat engine, the target function can be simplified by $\chi = P$, where $P$ is the power output. Optimization then simply turns into finding the EMP, a very meaningful parameter for a heat engine. From Eqs. (7-9) and (11), the power output equations for $k_r$, $k_x$ and $k_y$ filtered heat engines can be written, respectively, as:

$$P^r = \frac{2\pi \Gamma_r k_B T_H \sqrt{2m^* E_r^p}}{h^2} \left[\lambda_H^r - (1-\eta_C)\lambda_C^r\right]\left(\frac{1}{1+e^{\lambda_H^r}} - \frac{1}{1+e^{\lambda_C^r}}\right) \quad (14a)$$

$$P^x = \frac{\pi \Gamma_x (k_B T_H)^{3/2} \sqrt{2\pi m^*}}{h^2}\left[\lambda_H^x - (1-\eta_C)\lambda_C^x\right]\left[\sqrt{1-\eta_C}\,\text{Li}_{1/2}\left(-e^{-\lambda_C^x}\right) - \text{Li}_{1/2}\left(-e^{-\lambda_H^x}\right)\right] \quad (14b)$$

$$P^y = \frac{\pi \Gamma_y (k_B T_H)^2 \sqrt{2m^*/E_y^p}}{h^2}\left[\lambda_H^y - (1-\eta_C)\lambda_C^y\right]\left[\log\left(1+e^{-\lambda_H^y}\right) - (1-\eta_C)\log\left(1+e^{-\lambda_C^y}\right)\right] \quad (14c)$$

where $\lambda_{H/C}^i = \left(E_i^p - \mu_{H/C}\right)/k_B T_{H/C}$ are dimensionless scaled energies. The optimizations can be produced by focusing on the electrochemical potentials $\mu_{H/C}$ and temperature ratio $\tau$ (or Carnot efficiency $\eta_C = 1 - \tau$), where $T_H$ and $E_i^p$ can be considered as arbitrary constants, and $\Gamma_i$ can also be regarded as a small constant. After setting the values of $T_H$, $E_i^p$ and $\Gamma_i$, the maximum power can then be obtained from $\partial P^i / \partial \lambda_H^i = \partial P^i / \partial \lambda_C^i = 0$. The EMP values can then be obtained through numerical computation, as shown in Fig. 4. The relationship $\lambda_C^i \geq \lambda_H^i$ can also be derived from the total entropy production rate of the system (refer to Eq. (12)). The corresponding numerical results are illustrated in Fig. 4 below.

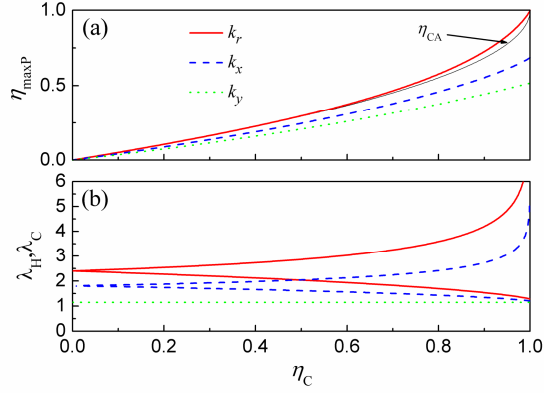

**Fig. 4.** (Color online) (a) Efficiencies at maximum power (EMP) with respect to $\eta_C$ for the $k_r$, $k_x$ and $k_y$ filtered heat engines. CA efficiency is just smaller than the EMP for $k_r$ filter, but much larger than those for $k_x$ and $k_y$ filters. (b) The corresponding dimensionless scaled energies, $\lambda_C^i$ (upper) and $\lambda_H^i$ (lower), respectively, with respect to $\eta_C$ for $k_r$ and $k_x$ filters, while $\lambda_C^y = \lambda_H^y = 1.14455$ for $k_y$ filter.

From these results, it is evident that the EMP for $k_r$ filtered heat engines is larger than for $k_x$ and $k_y$ alternatives, with the $k_y$ filter yielding lowest results. The CA efficiency is slightly less than the EMP for $k_r$ filter, but somewhat higher than the EMP for $k_x$ and $k_y$ filters. The EMP for $k_r$ filtered heat engines is equivalent to that of a 1D-0D-1D quantum dot device [15], which may be because the same Fermi-Dirac statistics and total scaled energy are used to independently model these two instances. All EMPs for three filters increase from zero to a maximum value when Carnot efficiency increases from 0 to 1. However, maximum efficiencies are different for each filter, i.e., 1, 0.686552 and 0.516548, respectively, for $k_r$, $k_x$ and $k_y$ filters. In the latter case, $\lambda_C^y = \lambda_H^y$ can be obtained by solving the formula: $\partial P^y / \partial \lambda_H^y = \partial P^y / \partial \lambda_C^y = 0$, and both scaled energies calculate to be 1.14455 at the maximum power, as shown in Fig. 4b. Even more interesting, the EMP in this instance can be approximated as: $\eta_{\max P}^{y} = \eta_{C} / (2.87186 - 0.93593\eta_{C})$, which is the same as the result for a 1D-system [35] where $\Gamma \to \infty$, since the energy level in the $k_y$ filter conductor tends to a single level when $\Gamma_y$ is small enough, causing the 2D-system to degenerate into 1D-system as $\Gamma \to \infty$, because of the half-freedom in $x$-direction (e.g., $k_x > 0$).

Similar results can be found in the refrigerator case. From $\partial \chi^{i} / \partial \lambda_{H}^{i} = \partial \chi^{i} / \partial \lambda_{C}^{i} = 0$, the COP at maximum $\chi$ (i.e. $\varepsilon_{\max \chi}$ in Fig. 5(a)) can be obtained by numerical computation. As expected, the $\varepsilon_{\max \chi}^{r}$ of $k_r$ filtered refrigerator is a little larger than the CA COP $\varepsilon_{CA}$, which is also larger than those of the $k_x$ and $k_y$ filtered refrigerators. The difference is that dimensionless scaled energies are observed to obey the relationship: $\lambda_{C}^{i} \leq \lambda_{H}^{i}$ (refer to Fig. 5(b)), which can also be derived by using $S^{i} \geq 0$. Moreover, it is observed that $\varepsilon_{\max \chi}^{x} > \varepsilon_{\max \chi}^{y}$ when $\tau < 0.972$. This result may be due to the cooling rate for $k_y$ filtered refrigerators being just large enough, compared to that for $k_x$ filtered refrigerators, when $\tau > 0.972$.

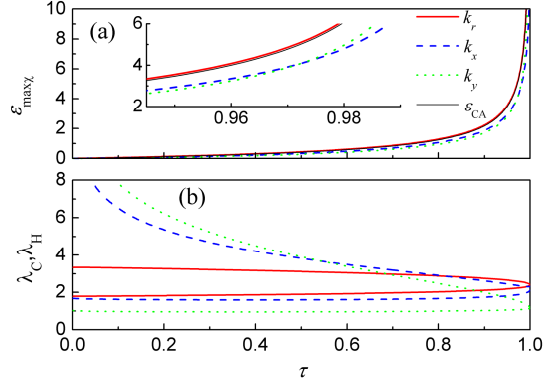

**Fig. 5.** (Color online) (a) COPs at maximum $\chi$ versus $\tau = T_{C}/T_{H}$ for $k_r$, $k_x$ and $k_y$ filtered refrigerators. The CA COP is slightly smaller than that for $k_r$ filter (just below the red line), but larger than for $k_x$ and $k_y$ filters. The COP at maximum $\chi$ for $k_x$ filter is slightly greater than that for $k_y$ filter up to around $\tau = 0.972$, and the trend then reverses. (b) The corresponding variation in dimensionless scaled energies, $\lambda_{H}^{i}$ (upper) and $\lambda_{C}^{i}$ (lower), respectively, for the three filters.

## C. $\Gamma_i$ is several multiples of $k_B T$

A more practical situation occurs when $\Gamma_i$ is several multiples of $k_B T$. Overall performance is bounded by the two limiting cases A and B just outlined. However, the case considered here is of interest because it comes closest to being the most feasible for practical thermoelectric devices, at least with current state-of-the-art capabilities. Much research has been carried out on performance optimization, especially for the heat engine. Taking the 1D heat engine with quantum dots as an example [36], both the maximum efficiency and EMP have been shown by others to decrease monotonically when $\Gamma$ increases; as expected, maximum efficiency is always larger than EMP. With increasing $\Gamma$, maximum power increases initially and then degrades slowly, and the peak is generally reached at around: $\Gamma \approx 2.25 k_B T$. As a point of differentiation to prior research, we have focused on exploring the performance optimization of the refrigerator, where $\Gamma_i$ is several multiples of $k_B T$.

Fig. 6 shows the optimization results for the maximum COP $\varepsilon_{max}^i$ with respect to $\Gamma_i$ for the three filtered refrigerators. Fig. 6b shows that all maximum COPs decrease monotonically from their initial values as $\Gamma_i$ increases from zero. The initial values at $\Gamma_i = 0$ are: $\varepsilon_C$, $0.807337\varepsilon_C$ and $0.7275\varepsilon_C$, respectively, for $k_r$, $k_x$ and $k_y$ filtered refrigerators (also shown in Fig. 3(a)). The corresponding energy peaks at the center of the resonance $E_i^p$, also decrease initially and then reverse this trend slightly as $\Gamma_i$ continues to increase. The initial values of $E_i^p$ are estimated to be: 1.18 eV, 1.1504 eV and 1.13095 eV, respectively, for $k_r$, $k_x$ and $k_y$ filtered refrigerators. All cooling rates at maximum COP increase initially and then decrease and finally vanish. This situation differs from that for the heat engine because of the finite cooling region. The $k_r$-filter is most beneficial in achieving higher $\varepsilon_{max}^i$ and cooling rate at maximum COP, followed by the $k_x$-filter, and finally the $k_y$-filter.

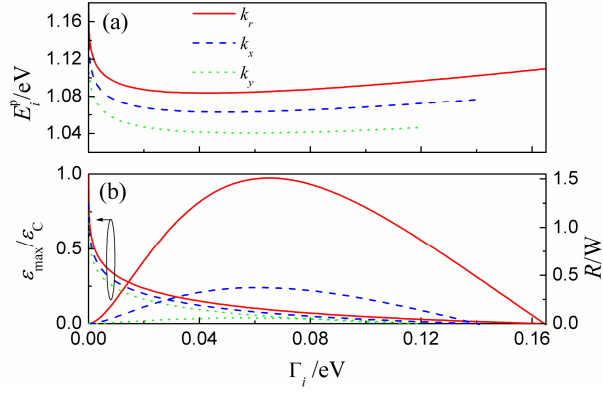

**Fig. 6.** (Color online) Maximum COP with respect to FWHM $\Gamma_i$ for $k_r$, $k_x$ and $k_y$ filtered refrigerators. (a) The corresponding peak energy at the center of the resonance $E_i^p$. (b) Relative maximum COP (left axis) and the corresponding cooling rate at maximum COP (right axis) with respect to $\Gamma_i$. Other parameter values are as stated in Fig. 3.

Likewise, in addition to the maximum COPs, the COPs at maximum cooling rate $\varepsilon_{\max R}^i$ are important performance parameters for finite $\Gamma_i$, and may even represent more meaningful parameters for practical applications. Fig. 7 shows the numerical results of $\varepsilon_{\max R}^i$ at the same given parameters as in Fig. 3. When $\Gamma_i$ increases gradually from zero, all maximum cooling rates of the three filtered refrigerator cases increase initially and finally decrease back to zero. Similarly, all $\varepsilon_{\max R}^i$ curves show a monotonic decrease. The corresponding energy peaks at center of the resonance $E_i^p$ also increase monotonously. Compared to $\varepsilon_{\max}^i$ in Fig. 6, the $\varepsilon_{\max R}^i$ are much smaller at low values of $\Gamma_i$, while they are comparable when $\Gamma_i$ increases. Even more interesting, the maximum cooling rates of the three cases peak around $\Gamma_i \approx 2k_B T$ (refer to Table 2), which corresponds with the result in reference [36] where the power of heat engine with quantum dots peaks at about $\Gamma \approx 2.25 k_B T$.

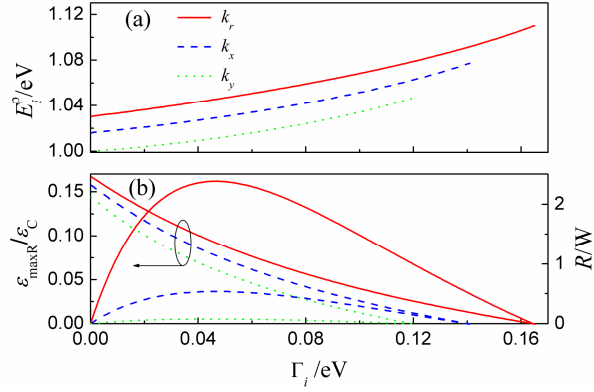

**Fig. 7.** (Color online) COP at maximum cooling rate with respect to FWHM $\Gamma_i$ for $k_r$, $k_x$ and $k_y$ filtered refrigerators. (a) The corresponding energy peak at center of the resonance $E_i^p$. (b) Relative COP at maximum cooling rate (left axis) and the corresponding maximum cooling rate (right axis), with respect to $\Gamma_i$. Other parameter values are as stated in Fig. 3.

Generally, the maximum efficiency or maximum COP can be predicted by the thermoelectric figure of merit $ZT$ from [37], as:

$$\eta_{\max} = \frac{M-1}{M+\tau}\eta_C \tag{15a}$$

$$\varepsilon_{\max} = \frac{M-1/\tau}{M+1}\varepsilon_C \tag{15b}$$

where $M = \sqrt{1+ZT}$. $ZT$ can be estimated from the calculated maximum efficiency or maximum COP.

We have chosen refrigerators as the example below. Based on the data in Fig. 6, $ZT$ decreases monotonically with the increasing FWHM for all three filtered refrigerator cases, as depicted in Fig. 8. The maximum values of $ZT$ for which $\Gamma_i \to 0$ are shown in Table 2, and are computed to be: $\infty$, 98.1559 and 44.5247, respectively, for $k_r$, $k_x$ and $k_y$ filtered refrigerators. Fig. 8 clearly shows that the $k_r$ filter provides the most beneficial thermoelectric figure of merit, followed by the $k_x$ filter, and then the $k_y$ filter. Narrower FWHM achieves higher $ZT$, however, suppresses the cooling rate. So the better strategy would be to choose a value for $\Gamma_i$ from zero to $2k_BT$, notwithstanding that $ZT$ is relatively small when $\Gamma_i \approx 2k_BT$ (refer to Table 2). Finally, it should be noted that the numerical calculations here do not include the impacts of heat loss, and calculated values of COP and $ZT$ are therefore optimistic.

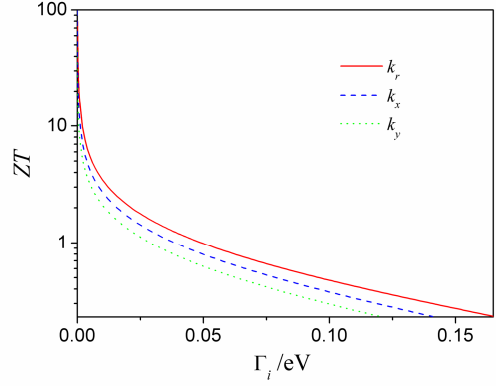

**Fig. 8.** (Color online) The estimated thermoelectric figure of merit $ZT$ with respect to FWHM $\Gamma_i$ for $k_r$, $k_x$ and $k_y$ filtered refrigerators. Other parameter values are as stated in Fig. 3.

**Table 2.** The thermoelectric figure of merit at maximum $\varepsilon_{\max}^i$ and the peak of maximum cooling rate, where FWHM at maximum cooling rate peak are $\Gamma_r^p = \Gamma_x^p = 0.047\,\text{eV}$ and $\Gamma_y^p = 0.045\,\text{eV}$. The Data are obtained from Figs. 6-8.

| FWHM | $\Gamma_i \to 0$ | | | $\Gamma_i \to \Gamma_i^p$ | | |
|---|---|---|---|---|---|---|
| Parameters | $\varepsilon_{\max R}^i/\varepsilon_C$ | $\varepsilon_{\max}^i/\varepsilon_C$ | $ZT$ | $\varepsilon_{\max R}^i/\varepsilon_C$ | $\varepsilon_{\max}^i/\varepsilon_C$ | $ZT$ |
| $k_r$ | 0.1675 | 1 | $\infty$ | 0.0917 | 0.1314 | 1.0465 |
| $k_x$ | 0.1579 | 0.8073 | 98.1559 | 0.0770 | 0.1046 | 0.8431 |
| $k_y$ | 0.1445 | 0.7275 | 44.5247 | 0.0632 | 0.0825 | 0.6922 |

## V. Conclusions

Theoretical models for three types of nanoscaled thermoelectric heat engines and refrigerators have been successfully developed and simulated through numerical computation, based on 2D electronic hot/cold reservoirs with interconnecting electron transport media acting, respectively, as momentum-dependent $k_x$, $k_y$ and $k_r$ filters. Important theoretical expressions for thermoelectric performance parameters have been derived, based on Fermi-Dirac electron distribution with resonant transport in the form of a Lorentzian resonance. These formulas have been used to numerically simulate and evaluate the performance of various device configurations as thermoelectric heat engines and refrigerators with 2D electron reservoirs.

When the FWHM of the Lorentzian resonance tends to zero, the upper bounds of several thermoelectric parameters have been identified. For heat engine operation, the theoretical maximum efficiency with respect to the resonance center $E_i^p$ and the EMP related to the temperature ratio $\tau$ were calculated. Similarly, the theoretical maximum COP and COP at maximum $\chi$ were also obtained for operation as refrigerators. The computed values for EMP and COP at maximum $\chi$ represent theoretical universal $\tau$-dependent bounds. For practical applications, refrigerators with finite FWHM were specifically chosen for study. In all three ($k_x$, $k_y$ and $k_r$) filtered refrigerator cases, a value of FWHM can be found in the range $< 2k_B T$ that achieves enhanced thermoelectric performance where both the cooling rate and thermoelectric figure of merit are considered, jointly. For all situations, it appears that the $k_r$ filter is able to achieve the best thermoelectric performance, followed by the $k_x$ filter, and then the $k_y$ filter. It is believed that these results may provide useful guidance in designing nanoscaled thermoelectric devices with 2D electron reservoirs, and a similar approach can apply to 3D electron systems under the parabolic approximation. However, the results cannot be adopted directly for situations in which electrons have linear (or other) dispersion relationships; this topic warrants further in-depth study.

**Acknowledgements**

X. Luo would like to thank Björn Sothmann for the meaningful discussion and advice. This work is supported by National Natural Science Foundation (No. 11365015), Program for New Century Excellent Talents in University of Ministry of Education of China (No. NCET-11-0096), the Fundamental Research Funds for the Central Universities and Research and Innovation Project for College Graduates of Jiangsu Province (No. CXZZ13_0081), People's Republic of China.